\documentclass[twocolumn,aps,showpacs,amsmath,amssymb,preprintnumbers,epsf]{revtex4}
\usepackage{graphicx}
\usepackage{dcolumn}
\usepackage{color}
\usepackage{bm}
\def\eqn{\end{equation}\noindent}
\def\eqnr{\end{eqnarray}\noindent}
\def\beqr{\begin{eqnarray}}
\def\beq{\begin{equation}}

\def\beq{\begin{equation}}

\def\beq{\begin{equation}}
\def\eqn{\end{equation}}
\begin {document}

\title{\textbf{Intercellular synchronization of diffusively coupled astrocytes}}
\author{Md. Jahoor Alam$^1$, Latika Bhayana$^1$, Gurumayum Reenaroy Devi$^1$,
Heisnam Dinachandra Singh$^1$, R.K. Brojen Singh$^1$, B. Indrajit Sharma$^2$.}
\affiliation{$^1$Centre for Interdisciplinary Research in Basic Sciences, Jamia Millia Islamia,
New Delhi 110025, India\\$^2$Department of Physics, Assam University, Silchar 788 011, Assam, India\\
}
\date{\today}

\begin{abstract}

{\begin{center}\bf ABSTRACT\end{center}}
{We examine the synchrony of the dynamics of localized $[Ca^{2+}]_i$ oscillations in internal pool of astrocytes via diffusing coupling of a network of such cells in a certain topology where cytosolic $Ca^{2+}$ and inositol 1,4,5-triphosphate (IP3) are coupling molecules; and possible long range interaction among the cells. Our numerical results claim that the cells exhibit fairly well coordinated behaviour through this coupling mechanism. It is also seen in the results that as the number of coupling molecular species is increased, the rate of synchrony is also increased correspondingly. Apart from the topology of the cells taken, as the number of coupled cells around any one of the cells in the system is increased, the cell process information faster.
}
\\\\
\emph KEYWORDS
{: Cell signaling, synchronization, diffusive coupling, network topology, chemical coupling}
\end{abstract}
\maketitle

\section{Introduction}

The astrocytes in the central nervous system have various important roles, namely, taking active part in signal processing \cite{cor,cha,dan}, interact with the neighbouring neurons \cite{pas,kan,new} etc. which leads to important responsibility of the cells in predicting disease states \cite{tia}. Since these cells interact with the environment, $Ca^{2+}$ waves propagated through the cells display oscillation in the nonpropagating internal stored $[Ca^{2+}]_i$ inside the cells \cite{boi}. These oscillations are sustained due to interaction of inositol 1,4,5-triphosphate (IP3) with extracellular, cytosolic and endoplasmic $Ca^{2+}$ through inositol cross coupling and calcium induced calcium release mechanisms \cite{mey,dupe}.

The coupling among a network of these astrocytes in a certain topology is done through the process of cell signaling. This cell signaling is being considered as a means of complex communication and information processing among individual astrocytes, astrocytes with neurons, governing basic cellular activities and to coordinate various actions involving various complex coupling mechanisms \cite{ast}. It has been predicted that $Ca^{2+}$ variation in IP3 concentration is necessary for the intercellular $Ca^{2+}$ waves propagation and this variation is initiated by IP3 diffusion through gap junctions to communicate neighbouring cells \cite{ley}. Therefore synchrony in the dynamics of the local $[Ca^{2+}]_i$ oscillations is believed to be due to chemical coupling i.e. due to exchange of cytosol $Ca^{2+}$ waves and IP3 \cite{ber,cal,kus,koi} and can exhibit synchronization of the cells over long distances \cite{kuc}. However, this idea of synchrony of $[Ca^{2+}]_i$ oscillation by chemical coupling is considered to be waek by claiming that these chemical wave propagate relatively slow and coupling through release of $Ca^{2+}$ is very weak as the effective diffusion of it is limited to very short distance \cite{imt}. They proposed electrochemical rather than chemical coupling, in which strong electrical coupling combined with weak chemical coupling, is responsible for the synchrony of these cells and is an effective means of long range signaling \cite{imt}. However, it has been predicted that in astrocytes intercellular $Ca^{2+}$ waves can travel over several hundred micrometers and may prodive a long range synchrony \cite{cor}.

The aim of this work is to try to raise this issue and try to look for a reasonable solution regarding synchrony of astrocytes via chemical coupling and possible long range interaction among the cells. We organize our work as following. We first study the model of $Ca^{2+}$ oscillations in astrocytes developed by Houart et. al. \cite{hou} in section II. We introduce possible diffusive coupling mechanisms among a topological network of coupled cells to see whether there is synchrony behaviours in the local $[Ca^{2+}]_i$ oscilations and look for long range communication among them. We present our simulation results based on the model discussed in section III and some conclusions based on our results are drawn in section IV.

\section{Materials and methods}

The basic single cell model which is described in Fig. 1 involves three key variables; the free calcium concentration in the cytosol (X), the concentration of stored $Ca^{+2}$ in the internal pool (Y), and the inositol 1,4,5-triphosphate, IP3 (Z) \cite{hou, dup, bor}. The time evolution of these three variables is governed by the following differential equations:
\begin{eqnarray}
\label{eq1}
\frac{dX}{dt}&=&V_0+V_1\beta-V_2+V_3+k_fY-kX\nonumber \\
\frac{dY}{dt}&=&V_2-V_3-k_fY\nonumber \\
\frac{dZ}{dt}&=&\beta V_4-V_5-\epsilon Z\\
V_2&=&V_{M2}\frac{X^2}{K_2^2+X^2}\nonumber \\
V_3&=&V_{M3}\frac{X^m}{K_X^m+X^m}\frac{Y^2}{K_Y^2+Y^2}\frac{Z^4}{K_Z^4+Z^4}\nonumber \\
V_5&=&V_{M5}\frac{Z^p}{K_5^p+Z^p}\frac{X^n}{K_d^n+X^n}\nonumber
\end{eqnarray}
where, $V_0$ indicates the constant input of $Ca^{2+}$ and $V_1$ returns to maximum rate of stimulus-induced influx of $Ca^{2+}$ from extracellular medium. The parameter $\beta$ is the degree of stimulation of the cell, $V_1$ and $V_2$ are the pumping and release of $Ca^{2+}$ from cytosol to internal store and internal store to cytosol respectively in the $Ca^{2+}$ induced $Ca^{2+}$ release (CICR) process with the maximum rates $V_{M2}$ and $V_{M3}$. $K_2$, $K_x$, $K_y$ and $K_z$ are the threshold values release, pumping, activation of release by $Ca^{2+}$ and IP3 (Z) respectively. $k_f$ is the passive, linear leak rate constant of X and Y; $k$ is the rate of $Ca^{2+}$ diffuse into extracellular medium; $V_4$ relates to rate of stimulus-induced synthesis of Z; $V_5$ is the phosphorylation rate of Z by the 3-kinase; $V_{M2}$ is the maximum value and half saturation constant $K_5$; $K_d$ corresponds to the threshold $Ca^{2+}$ level; $\epsilon$ is the reflected Z to mobilize in $Ca^{2+}$; $m$, $n$ and $p$ are Hill co-efficients.
\begin{figure}
\label{fig11}
\begin{center}
\includegraphics[height=180 pt]{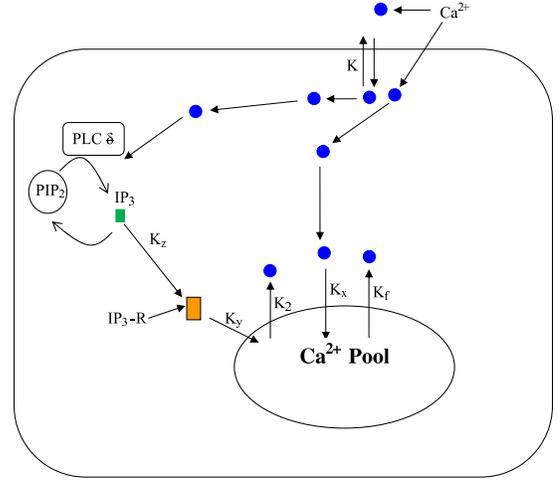}
\caption{(A) The schematic diagram of reaction network of molecular mechanisms of the model of $Ca^{2+}$ of the model.}
\end{center}
\end{figure}
\begin{figure}
\label{fig12}
\begin{center}
\includegraphics[height=150 pt]{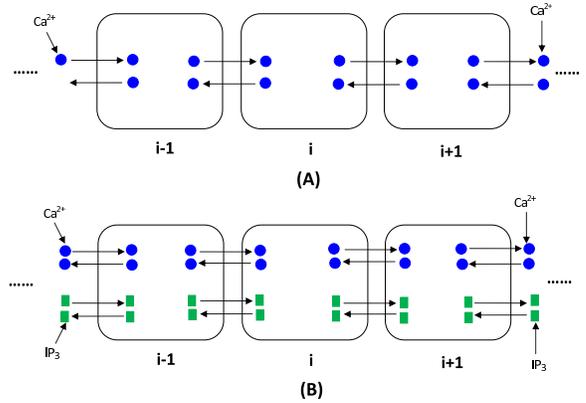}
\caption{(A) The schematic diagram of diffusive coupling of a chain of cells via (A) Diffusing molecule, $Ca^{2+}.$, (B) Diffusing molecules, $Ca^{2+}.$ and IP3.}
\end{center}
\end{figure}

The oscillations in in the variables X, Y and Z exhibit various types, namely simple oscillation, bursting, chaotic and quasiperiodic subject to different values of reaction constants and parameters in the set of equations \ref{eq1} \cite{hou,zhu}. These complex oscillations of the variables are in fact significantly stimulated by the variation of populations of species IP3 (Z) and stored $Ca^{2+}$ in the internal pool (Y) supported by some experimental reports\cite{bor,cao}.
\begin{figure}
\label{fig2}
\begin{center}
\includegraphics[height=170 pt]{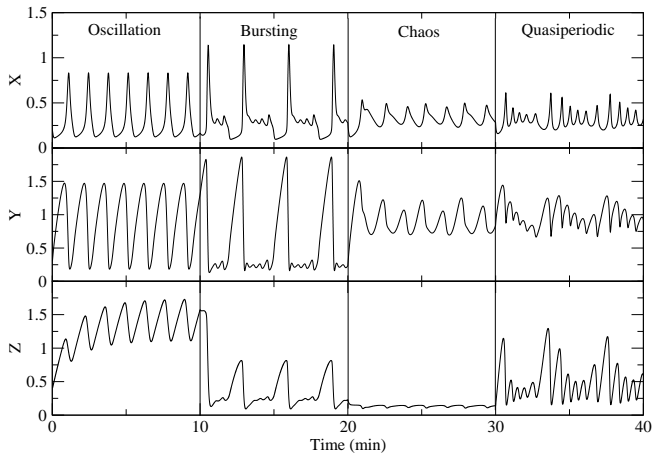}
\caption{The plot showing the time evolution of concentration of X, Y and Z showing (i) simple oscillation for the parameters \cite{hou}: $V_0$=2, $V_1=2$, $\beta=0.5$, $V_{M2}=6$, $k_2=0.1$, $V_{M3}=20$, $K_x=0.5$, $K_y=0.2$, $K_z=0.2$, $V_{M5}=5$, $k_5=1$, $k_d=0.4$, $k_f=1$, $k=10$, $\epsilon=0.1$, $V_4=2$, $m=2$, $p=2$ and $n=4$. (ii) bursting for the parameters: $V_0=2$, $V_1=2$, $\beta=0.46$, $V_{M2}=6$, $k_2=0.1$, $V_{M3}=20$, $Kxz=0.3$, $K_y=0.2$, $K_z=0.1$, $V_{M5}=30$, $k_5=1$, $k_d=0.6$, $k_f=1$, $k=10$, $\epsilon=0.1$, $V_4=2.5$, $m=4$, $p=1$ and $n=2$. (iii) chaos for the parameters: $V_0=2$, $V_1=2$, $\beta=0.65$, $V_{M2}=6$, $k_2=0.1$, $V_{M3}=30$, $K_x=0.6$, $K_y=0.3$, $K_z=0.1$, $V_{M5}=50$, $k_5=0.3194$, $k_d=1$, $k_f=1$, $k=10$, $\epsilon==13$, $V_4=3$, m=2, p=1 and $n=4$ and (iv) quasiperiodicity for the parameters: $V_0=2$, $V_1=2$, $\beta=0.51$, $V_{M2}=6$, $k_2=0.1$, $V_{M3}=20$, $K_x=0.5$, $K_y=0.2$, $K_z=0.2$, $V_{M5}=30$, $k_5=0.3$, $k_d=0.5$, $k_f=1$, $k=10$, $\epsilon==0.1$, $V_4=5$, m=2, p=2 and $n=4$}
\end{center}
\end{figure}
\begin{figure*}
\label{fig3}
\begin{center}
\includegraphics[height=350 pt]{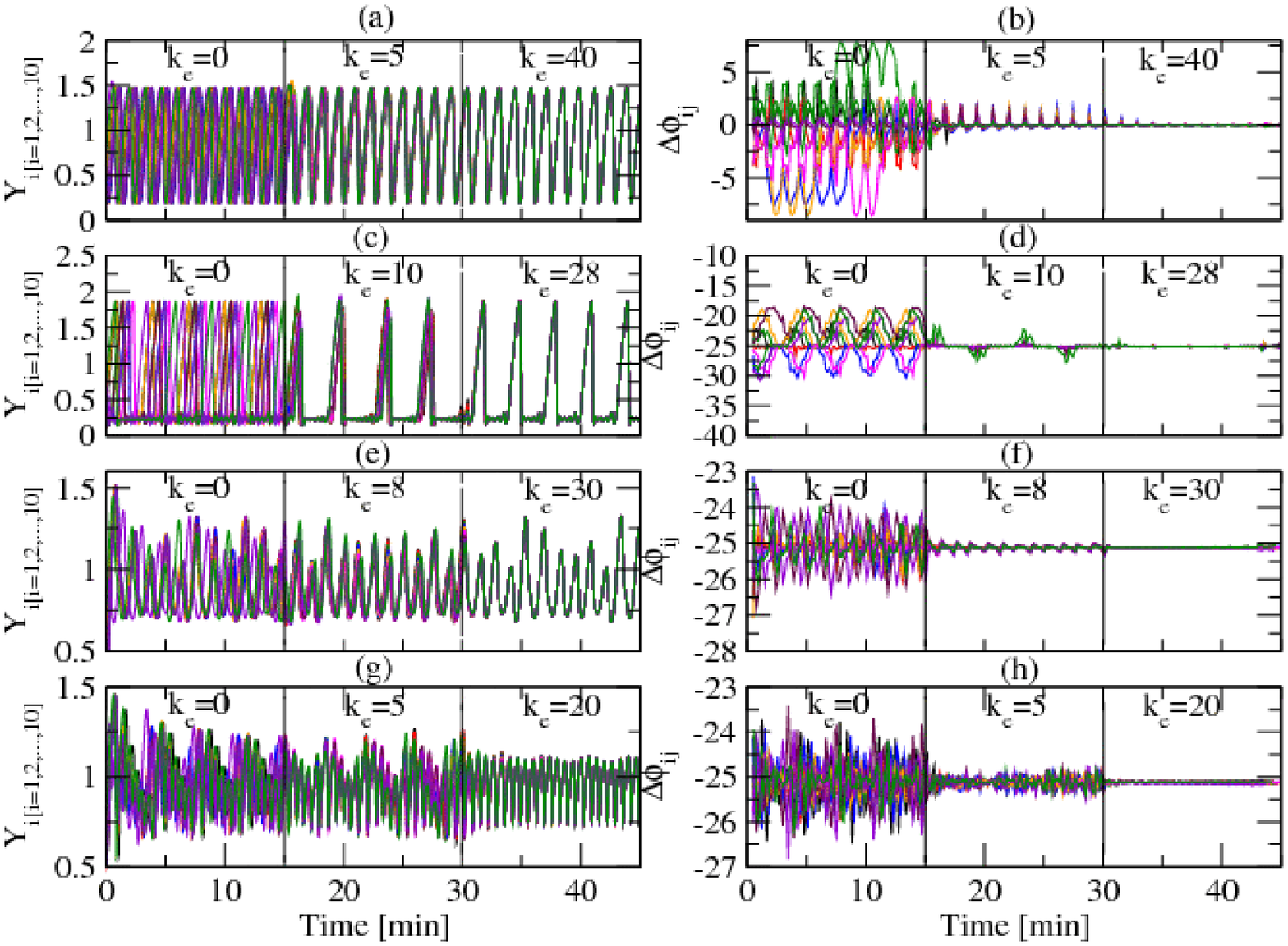}
\caption{Plots of $Y$ (left panels) and phase difference $\Delta\phi$ (right panels) as a function of time for different values of $k_e$: (i) simple oscillation in (a) and (b) for $k_e$=0, 5 and 40; (ii) bursting in (c) and (d) for $k_e$=0, 10 and 28; (iii) chaos in (e) and (f) for $k_e$=0, 8 and 30; (iv) quasiperiodic in (g) and (h) for $k_e$=0, 5 and 20; for 10 cells out of 50 cells showing desynchronized, weakly synchronized and strongly synchronized regimes.}
\end{center}
\end{figure*}

Consider a group or network of $N$ such identical cells which are coupled by the exchange of free cytocelic $Ca^{2+}$ ions (X) and IP3 (Z) by diffusing through the ion-channels on each cell surface. Among these cells, the information of individual cell reactions is transmitted via these diffusing molecules to each neighbouring cells in the network and a unique globally co-ordinated behaviour of the localized molecules i.e. $[Ca^{2+}]_i$ stored in internal pool (Y) of each individual cells in the network will be exhibited. If we consider a simple topological network of these cells diffusively coupled in series in the form a chain as shown in Fig. 2 (A) and (B); we have the following two mechanisms of diffusing coupling in the topology; (i) {\bf Single molecule diffusive coupling:}  In this case of single molecular species coupling mechanism, where $X$ being taken as coupling molecule, the species $X_{i}$ and $X_{i\pm 1}$ can interconvert via additional reaction channels, $X_{i}\stackrel{k_{x1}}\rightarrow X_{i-1}$, $X_{i-1}\stackrel{k_{x2}}\rightarrow X_i$, $X_{i}\stackrel{k_{x3}}\rightarrow X_{i+1}$, $X_{i+1}\stackrel{k_{x4}}\rightarrow X_i$, where the diffusing rates in all these additional reactions are $k_{x1}$, $k_{x2}$, $k_{x3}$ and $k_{x4}$. Deterministically this corresponds to two bidirectional diffusive couplings i.e. [$k_{x1}(X_i-X_{i-1})$ and $k_{x2}(X_{i-1}-X_i)$] and [$k_{x3}(X_i-X_{i+1})$ and $k_{x4}(X_{i+1}-X_i)$] are respectively incorporated bidirectionally. So the synchronization in other variables $Y_k,(k=1,2,...,N)$ occurs when the rates $k_{x1}$, $k_{x2}$, $k_{x3}$ and $k_{x4}$ are sufficiently large. Now, taking $k_{x1}=k_{x2}=k_{x3}=k_{x4}=k_x$ for simplicity, we have the following diffusively coupled differential equations of the cells in the network topology we considered,
\begin{eqnarray}
\label{eq2}
\frac{dX_i}{dt}&=&F_i+\sum_{j=0}^{1}k_x(X_{i+2j-1}-X_i)\nonumber \\
\frac{dY_i}{dt}&=&G_i\nonumber \\
\frac{dZ_i}{dt}&=&H_i
\end{eqnarray}
and (ii) {\bf Global diffusive coupling:} In this case, two or more molecular species are considered as diffusively coupling molecules in the cells in the network. For every diffusing molecular species, four additional reaction channels are to be added with different rates. In the model we considered, we have two diffusively coupling molecules i.e. $X$ and $Z$. Therefore we will have eight additional reaction channels; four for $X$ and four for $Z$ respectively. Taking the diffusive rates of each molecular species to be the same, the coupled differential equations are given by,
\begin{eqnarray}
\label{eq3}
\frac{dX_i}{dt}&=&F_i+\sum_{j=0}^{1}k_x(X_{i+2j-1}-X_i)\nonumber \\
\frac{dY_i}{dt}&=&G_i\nonumber \\
\frac{dZ_i}{dt}&=&H_i+\sum_{j=0}^{1}k_z(Z_{i+2j-1}-Z_i)
\end{eqnarray}
where, $X_i$, $Y_i$ and $Z_i$ are the variables of the ith cell in the chain of cells we considered. The functions in the above equations are defined by, $F_i=V_0+V_1\beta-V_2+V_3+k_fY_i-kX_i$, $G_i=V_2-V_3-k_fY_i$ and $H_i=\beta V_4-V_5-\epsilon Z_i$. The parameters $k_x$ and $k_z$ are the coupling constants of $X$ and $Z$ molecular species and are not necessary to have the same value. For the topology of the finite chain of cells, the coupling terms containing $X_0$, $Z_0$ in the first cell and $X_{N+1}$, $Z_{N+1}$ in the Nth cell are neglected since these molecular species are not in the domain of the system we considered. However, if the chain become a ring by connecting the two ends, then we have to apply the boundary conditions i.e. $X_0=X_N$, $Z_0=Z_N$ and $X_{N+1}=X_1$, $Z_{N+1}=Z_1$.

The measure of synchronization of the time evolution of two independent and identical systems can be possible \cite{sak} by defining an instantaneous phase for an arbitrary signal $\eta(t)$ via the Hilbert transform \cite{ros}
\begin{equation}
\label{eq4}
\tilde \eta (t)=\frac{1}{\pi}P. V. \int_{-\infty}^{+\infty}\frac{\eta (t)}{t-\tau}d\tau
\eqn
where $P. V.$ denotes the Cauchy principal value. The instantaneous phase $\phi(t)$ and amplitude $A(t)$ of a given arbitrary signal can be obtained through the relation, $A(t)e^{i\phi(t)}= \eta (t)+i\tilde \eta (t)$. For any given pair of signals $[(i,j);i,j=1,2,...,N, i\ne j]$, one can therefore obtain the instantaneous phases $\phi_{i}$ and $\phi_{j}$;  phase synchronization is then the condition that $\Delta\phi=m\phi_i-n\phi_j$ is constant with $m$ and $n$ being integers. Starting with different initial configurations, the temporal dynamics of the uncoupled oscillators will be  uncorrelated; however upon coupling, the dynamics can show phase synchrony \cite{pik,ros,nan}.

Another way to measure the rate of synchrony of two coupled oscillators is to plot the two corresponding variables $x,x^\prime$ from the two oscillators along the two axes of the two dimensional cartesian plane (Pecora-caroll type) \cite{pec}. If the oscillators are uncoupled then the points in the plane scattered away from the diagonal. However, if the oscillators are coupled then the points concentrated towards the diagonal. The rate of concentration of the points towards the diagonal measures the rate of synchrony.
\begin{figure*}
\label{fig4}
\begin{center}
\includegraphics[height=350 pt]{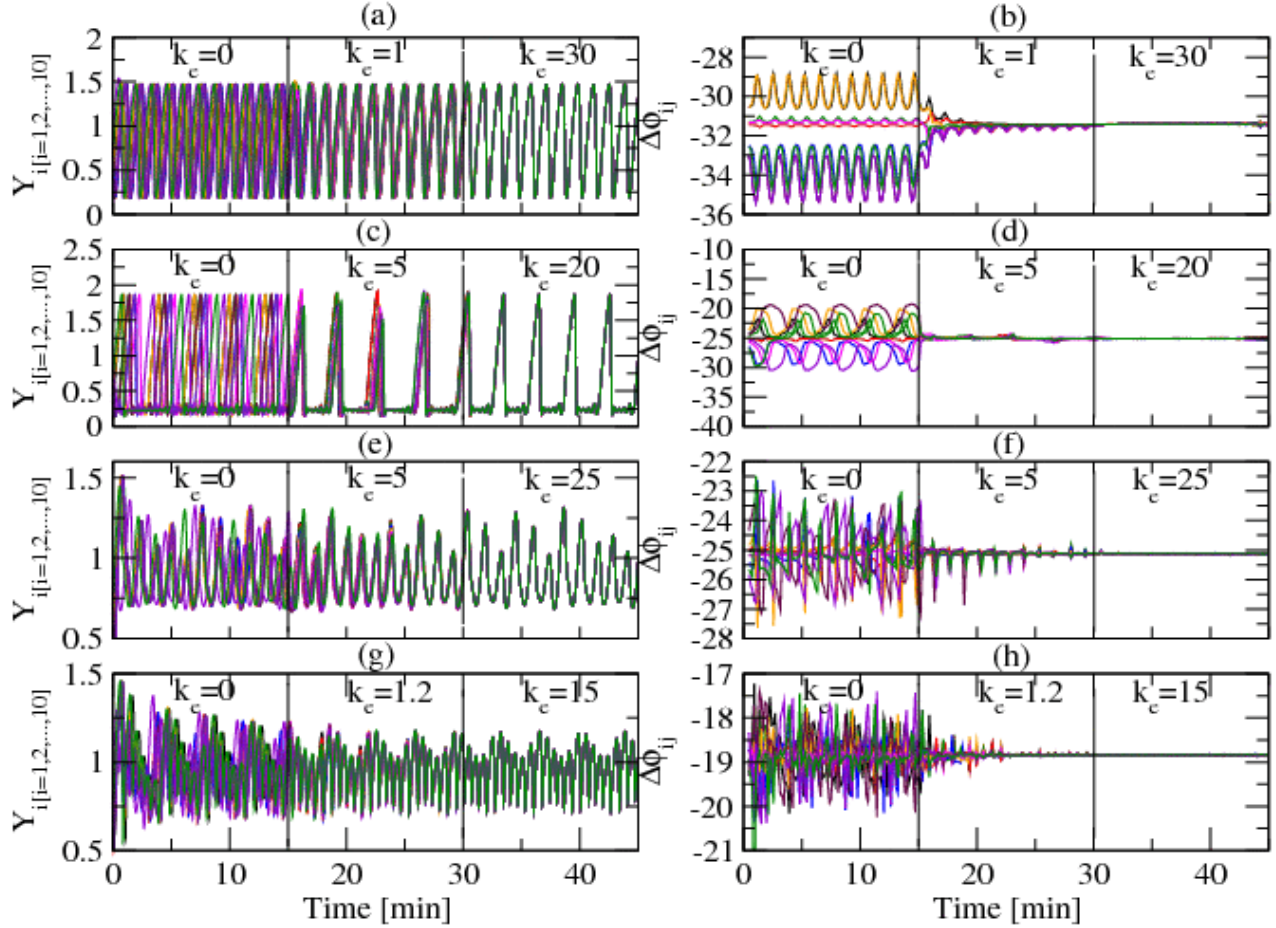}
\caption{Similar plots of $Y$ and $\Delta\phi$ (right panels) as a function of time for different values of $k_e$ for simple oscillations, bursting, chaos and quasiperiodic cases.}
\end{center}
\end{figure*}

\section{Results}

We present first one time temporal dynamics of the concentrations of the variables of single cell by standard numerical integration technique \cite{pre} of the set of differential equations (\ref{eq1}) for various values of four sets of parameters \cite{hou} given in the figure captions of Fig. 3 showing simple oscillation, bursting, chaos and quasiperiodicity. The time is measured in minutes. Then we take $N=50$ identical cells, diffusive coupling is employed with $X$ as diffusing coupler molecule at different rate constants at different times and solved the sets of coupled differential equations (\ref{eq2}) as shown in Fig. 4. The results of {\bf simple oscillations} are shown in Fig.~4 (a) and (b); the figure (a) shows plots of the dynamics of concentrations of the variable $Y_i(i=1,2,...,N)$ for the first $N=10$ cells and figure (b) shows the corresponding phase plot ($\Delta\phi$ verses time) for different rate constants i.e. $k_e=0$ (uncoupled), $k_e=5$ and $k_e=40$ switched on in time intervals (0-15), (15-30) and (30-45) minutes respectively. The uncoupled behaviours of the cells are shown during time interval (0-15) minutes proved by random fluctuation of $\Delta\phi$ with time during this time interval and pecora-caroll plot of $Y_1$ verses $Y_{10}$ in Fig. 6 (a) where the points spread randomly away from the diagonal. During time interval (15-30) minutes the behaviours of the cells start co-ordinating but is still weak indicating weak synchronization which is again proved by phase plot where the fluctuation is reduced enormousely and by pecora-caroll plot whose points start concentrated enormousely along the diagonal. However, during time interval (30-45), the cells exhibit strong co-ordinated nature i.e. synchronized state proved by the negligible fluctuation in the phase plot and diagonally aligned points in the pecora-caroll plot. We also noticed from the results that the dynamics of the variables of the cells do not synchronized immediately but takes some time after the coupling is switched on.
\begin{figure}
\label{fig5}
\begin{center}
\includegraphics[height=200 pt]{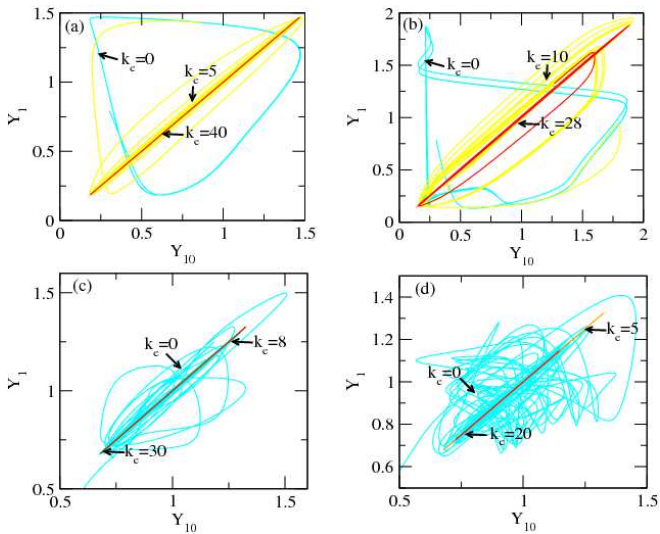}
\caption{The pecora-caroll type plot of variable Ys of diffusively coupled first and 10th cells for (a) oscillation, (b) bursting, (c) chaos and (d) quasiperiodic.}
\end{center}
\end{figure}

Similarly, for the {\bf bursting} case, the curves in Fig. 4 (c) and (d); and Fig. 6 (b) present how the dynamics of the variable $Y$ start co-ordinating their behaviours from uncoupled to weak, then to strong showing desynchronized, weak and strong synchronization. As explained above, the claim is based on the fluctuation rates in the phase plot (Fig 4. (d)) and spreading rate of the points from the diagonal in pecora-caroll plot (Fig. 6 (b)). The diffusing rates for desynchronization, weak and strong synchronization in this case are found to be $k_e=0, 10$ and 28. In the same way, the results for {\bf chaos} and {\bf quasiperiodic} are shown in Fig.4 (e) and (f) with Fig. 6 (c) (for {\bf chaos}) and Fig. 4 (g) and (h) with Fig. 6 (d) (for {\bf quasiperiodic}) respectively. The diffusing rates are found to be different as $k_e=0,8,30$ and $k_e=0,5,20$ for {\bf chaos} and {\bf quasiperiodic} respectively. 

Now the simulation results for the cases of simple oscillation, bursting, chaos and quasiperiodic, when two diffusively coupling molecular species namely $X$ and $Z$ are considered in the same topology of the cells, are shown in Fig. 5 [(a), (b)], [(c), (d)], [(e), (f)] and [(g), (h)] with Fig. 7 (a), (b), (c) and (d) respectively. The results are obtained by solving the set of coupled differential equations (\ref{eq3}). Similar behaviours of desynchronized, weakly synchronized and strongly synchronized are found as in the case where single molecular species diffusive coupling is employed. However, interestingly the mentioned behaviours are found in significantly lower diffusing rates i.e. $k_e=0, 1, 30$ for {\bf simple oscillations}, $k_e=0, 5, 20$ for {\bf bursting}, $k_e=0, 5, 25$ for {\bf chaos} and $k_e=0, 1.2, 15$ for {\bf quasiperiodic} respectively.

The results interestingly give the evidence that as the number of coupling molecular species are increased, the rate of synchronization also increases significantly. In other words, as the number of information carrying molecules (same molecular species) or(and) molecular species increases, the rate of the information processed by the cells also increases accordingly and the rate of correlation of the cells becomes stronger. In this coupling scheme,
\begin{eqnarray}
\label{eq5}
\frac{dP_i}{dt}&=&\Gamma_i+\sum_{j=0}^{1}k_j[P_{i+2j-1}-P_i]\nonumber \\
&&~~~~+\sum_{r=1}^Mk_r[P_r-P_i]
\end{eqnarray}
where, $\Gamma_i=\Gamma_i(X_i,Y_i,Z_i)$ is a function and $P_i=(X_i,Y_i,Z_i)$. The extra sum indicates extra reaction channels due to diffusively coupling of other $M$ cells indicated by cell number index, $r=1,2,...,M$, apart from the chain of cells. This extra term with increasing $M$ will significantly contribute to the increase of information transfer to the cell itself in ith position in the network topology from the coupling cells and to more other cells to show synchrony of more cells.

\section{Conclusion}

The diffusion of inositol 1,4,5-triphosphate and cytosolic calcium ion from one cell to another through the ion-channels on the surfaces of the cells couple the cells giving rise co-ordinated behaviour of the cells. Moreover, as the number of diffusing molecules among the cells increases, the amount of information transfer among the cells is also increased and the cells synchronized faster. If the number of diffusing channels is increased due to increase in neighbouring cells, then more cells will be coupled and information transfer is quicker. So we claim that the role of chemical coupling has significantly important role in the synchrony of astrocytes and long range information transfer among them.
 
\section{Acknowledgments}

This work is financially supported by UGC and carried out in center for Interdesciplinary research in basic sciences, Jamia Millia Islamia,New Delhi,India.


\begin{thebibliography}{99}

\bibitem{cor} A.H. Cornell-Bell, S.M. Finkbeiner, M.S. Cooper and S.J. Smith, Science {\bf 247}, 470-473 (1990).
\bibitem{cha} A.C. Charles, J.E. Merril, E.R. Ditksen and M.J. Sanderson, Neuron {\bf 6}, 983-992 (1991).
\bibitem{dan} J.W. Dani, A. Chernavsky and S.J. Smith, Neuron {\bf 8}, 429-440 (1992).
\bibitem{pas} L. Pasti, A. Volterra, T. Pozzan and G. Carmignoto, J. Neurosci. {\bf 17}, 7817-7830 (1997).
\bibitem{kan} J. Kang, L. Jiang, S.A. Goldman and M. Nedergaard, Nat. Neurosci {\bf 1}, 683-692 (1998).
\bibitem{new} E.A. Newman and K.R. Zahs, J. Neurosci. {18}, 4022-4028 (1998).
\bibitem{tia} G.F. Tian, H. Azmi, T. Takano, Q. Xu, W. Peng, J. Lin, N. Oberheim, N. Lou, X. Wang, H.R. Zielke, J. Kang and M. Nedergaard, Nat. Med. {\bf 11}, 972-981 (2005).
\bibitem{boi} S. Boitano, E.R. Dirksen and M.J. Sanderson, Science {\bf 258}, 292-295 (1992).
\bibitem{mey} T. Meyer and M. Stryer, Proc. Natl. Acad. Sci. {\bf 85}, 5051-5055 (1988), Annu. Rev. Biophys. Chem. {\bf 20}, 153-174 (1991).
\bibitem{dupe} G. Dupent and A. Goldbeter, Cell Calcium {\bf 14}, 311-322 (1993).
\bibitem{ast} A.R. Asthagiri and D.A. Lauffenburger, Annu. Rev. Biomed. Eng., 2, 31-53 (2000).
\bibitem{ley} L. Leybaert, K. Paemeleire, A. Strahonja and M.J. Sanderson, GLIA {\bf 24}, 398-407 (1998).
\bibitem{ber} M.J. Berridge, Nature {\bf 361}, 315-325 (1993).
\bibitem{cal} N. Callamaras, S.J. Marchant, X.P. Sun and I. Parker, J. Physiol. {\bf 509}, 8191 (1998).
\bibitem{kus} J.M. Kusters, W.P. van Meerwijk, D.L. Ypey, A.P. Theuvenet and C.C. Gielen, Am. J. Physiol. Cell Physiol. {\bf 294}, C917-930 (2008).
\bibitem{koi} S. Koizumi, FEBS J. {\bf 177}, 286-292 (2010). 
\bibitem{imt} M.S. Imtiaz, PY. von der Weid and D.F. van Helden, FEBS J. {\bf 277}, 278-285 (2010).
\bibitem{kuc} K.V. Kuchibhotla, C.R. Lattarulo, B.T. Hyman and B.J. Baeskai, Cell Calcium {\bf 14}, 711-723 (1993).
\begin{figure}
\label{fig6}
\begin{center}
\includegraphics[height=200 pt]{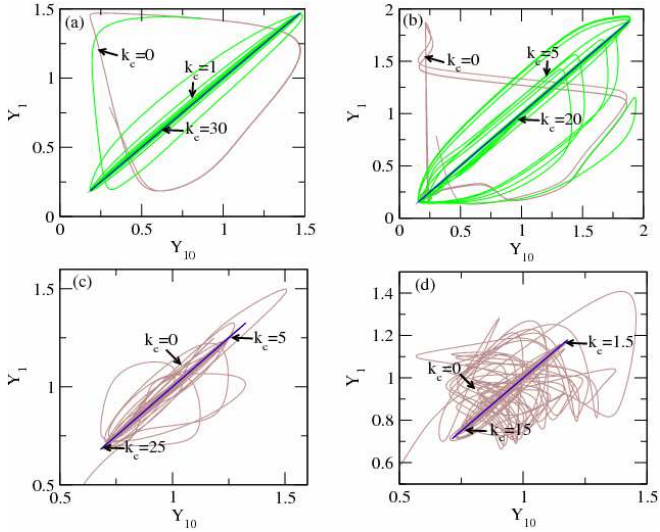}
\caption{Similar plots of Ys for (a) oscillation, (b) bursting, (c) chaos and (d) quasiperiodic.}
\end{center}
\end{figure}
\bibitem{hou} G. Houart, G. Dupont and A. Goldbeter, Bull. Math. Biol. {\bf 61}, 507-530 (1999).
\bibitem{dup} G. Dupont and A. Goldbeter, Cell Calcium {\bf 14}, 311-322 (1993).
\bibitem{bor} J.A.M. Borghans, G. Dupont and A. Goldbeter, Biophys. Chem. {\bf 66}, 25-41 (1997).
\bibitem{zhu} C.Z. Zhu, Y. Jia, L.Y. Yang and X. Zhan, Biophys. Chem. {\bf 125}, 201-212 (2007).
\bibitem{cao} D. Cao, G. Lin, E.M. Westphale, E.C. Beyer and T.H. Steinberg, J. Cell Sc. {\bf 110}, 497-504 (1997).
\bibitem{sak} H. Sakaguchi and Y. Kuramoto, Prog. Theor. Phys., {\bf  76}, 576 (1986).
\bibitem{pik} A. Pikovsky, M. Rosenblum and J. Kurths, {\it Synchronization: A  Universal Concept in Nonlinear Science} (Cambridge University Press, Cambridge, 2001).
\bibitem{ros} M. G. Rosenblum, A. S. Pikovsky and J. Kurths, Phys. Rev. Lett.,
{\bf  76}, 1804 (1996).
\bibitem{nan} A. Nandi, Santhosh G., R. K. B. Singh and R. Ramaswamy, Phys. Rev. E,
{\bf 76}, 041136 (2007).
\bibitem{pec} L. M. Pecora and T. L. Caroll, Phys. Rev. Lett., {\bf  64}, 821 (1990).
\bibitem{pre} W.H. Press, S.A. Teukolsky, W.T. Vetterling and B.P. Flannery, Numerical Recipe in Fortran, Cambridge University Press, 1992. 
\end{thebibliography}
\end{document}